\documentclass[iop,apjl]{emulateapj} 

\usepackage[usenames]{color}

\usepackage{soul}
\usepackage{color}

\slugcomment{Accepted to ApJL. Manuscript  LET30652 }

\newcommand{\blue}{\textcolor{blue}}
\newcommand{\kms}{km\,s$^{-1}$}

\newcommand{\teff}{$T_{\rm eff}$ }

\newcommand{\msun}{$\mathrm{M_{\odot}}$}

\def\ltsim{\raise 1.0pt \hbox {$<$} \kern-1.0em \lower 4.0pt \hbox {$\sim$}}

\shorttitle{An evolved solar twin revealed by CoRoT}

\shortauthors{do Nascimento et al.}

\begin{document}

\title{The future of the Sun: an evolved solar twin revealed by CoRoT}

\author{J.-D.~do Nascimento Jr.$^1$, Y.~Takeda$^2$, J.~Mel\'endez$^3$, J.S.~da Costa$^1$, G.F.~Porto de Mello$^{4}$ 
and M.~Castro$^1$}


\altaffiltext{1}{Universidade Federal do Rio Grande do Norte, UFRN, Dep. de F\'{\i}sica Te\'orica e Experimental,  
DFTE, CP 1641, 59072-970, Natal, RN, Brazil; dias@dfte.ufrn.br}
\altaffiltext{2}{National Astronomical Observatory of Japan 2-21-1 Osawa, Mitaka, Tokyo 181-8588}
\altaffiltext{3}{Departamento de Astronomia do IAG/USP, Universidade de S\~ao Paulo, Rua do Mat\~ao 1226, S\~ao Paulo, 05508-900, SP, Brazil}
\altaffiltext{4}{Observat—rio do Valongo, UFRJ, Ladeira do Pedro Ant\^onio, 43 20080-090 Rio de Janeiro, RJ, Brazil}


\begin{abstract}
The question of whether the Sun is peculiar within the class of solar-type stars has been the 
subject of active investigation over the past three decades. Although several solar twins 
have been found with stellar parameters similar to those of the Sun (albeit in a range of 
Li abundances and with somewhat different compositions), their rotation periods are 
unknown, except for 18 Sco, which is younger than the Sun and with a rotation period shorter 
than solar. It is difficult to obtain rotation periods for stars of solar age from ground-based 
observations, as a low activity level imply a shallow rotational modulation of their 
light curves. CoRoT has provided space-based 
long time series from which the rotation periods of solar twins as old as the Sun could be 
estimated. Based on high S/N high resolution spectroscopic observations gathered at the 
Subaru Telescope, we  show that the star CoRoT~ID~102684698 is a somewhat evolved solar 
twin with a low Li abundance. Its rotation period is 29 $\pm$ 5 days, 
compatible with its age (6.7 Gyr) and low lithium content  
$A_{\rm Li}$ \ltsim~0.85  dex. Interestingly, 
our CoRoT solar twin seems to have enhanced abundances of the refractory elements with 
respect to the Sun, a typical characteristic of most nearby twins. 
With a magnitude V $\simeq$ 14.1, ~ID~102684698 is  the first solar twin revealed by CoRoT,
the farthest field  solar twin so far known,  and the only solar twin older than the Sun for which a 
rotation period has been determined.
\end{abstract}

\keywords{stars: fundamental parameters --- stars: abundances ---  stars: rotation ---  stars: evolution --- Sun: fundamental parameters}


\section{Introduction}
\label{intro}
Modern stellar astrophysics is producing an amount of data never seen before. Two space missions, 
CoRoT \blue{{\citep{baglin06}}} and Kepler \blue{{\citep{Borucki}}}, are providing precise light curves 
observations for thousands of main-sequence stars, from which rotation periods  $P_{\mathrm {rot}}$ and 
solar-like oscillations can be studied in detail. CoRoT observes towards the  intersection between 
the equator and the Galactic plane and has identified thousands of solar-type dwarf stars. 
Among them, there are stars with fundamental  parameters very similar to the Sun, the so-called solar twins.
With CoRoT we can estimate periodic stellar variability in the range of 1 -- 50 days, a modulation 
that is a signature of the presence of spots on the star's surface. 

\vspace{0.3cm}

Although recent works have greatly expanded the number of solar twins and studied in detail 
their physical parameters and chemical abundances, their $P_{\mathrm {rot}}$ are mostly unknown, 
except for the solar twin 18 Sco \blue{\citep{portomello1997}}, which seems to have physical 
characteristics similar to solar \blue{\citep{bazot}}, a Li abundance about 
three times solar \blue{\citep{melendez2007,takeda2009}}, a younger age \blue{\citep{baumann}} 
and $P_{\mathrm {rot}}$  somewhat faster  than the Sun \blue{\citep{frick,petit}}
and a Sun-like activity cycle with a  shorter length  \blue{\citep{hall2007}}.

Besides the astrophysical importance of solar twins to assess to which point the Sun 
could be considered as a ``typical'' solar-type star 
\blue{{\citep{gustafsson1998, gusta2008}}},
solar twins are also important to calibrate fundamental relations between colors and temperature 
\blue{{\citep{portomello1997, melendez2010, ivan2012, luca2012}}}, and to test
non-standard stellar evolution models \blue{\citep[e.g.,][]{Nascimento_2009,castro2011}}.
A bonafide sample of solar twins with determined $P_{\mathrm {rot}}$ are also important to study the 
``Sun in Time''  (see   \blue{\citealt{dorren_guinan1994}}, \blue{\citealt{ribas2010}}), i.e., the evolution of the Sun 
through solar twins covering a range of ages.

\vspace{0.3cm}

Despite the fact that until 1997 only one solar twin was known \blue{\citep{portomello1997}}, 
the search for solar twins \blue{\citep{hardorp1978, cayrel1996, soubiran}}
has been greatly expanded in the last few years, and currently  more than two dozen solar twins 
are known \blue{\citep[e.g.,][]{melendez2006,takeda2007,melendez2007,melendez2009,ivan2009,datson}}. 
However, only for 18 Sco, a solar twin  younger than the Sun, its $P_{\mathrm {rot}}$ has been
determined \blue{\citep{frick,petit}}. 
The discovery of mature solar twins, with ages similar to the Sun's or higher, 
is highly desirable in order to study the rotational evolution of the Sun.

\begin{figure}   
\centering
\vspace{-0.5cm}
\hspace{-0.5cm}
\includegraphics[angle=0,width=9.5cm,height=15cm]{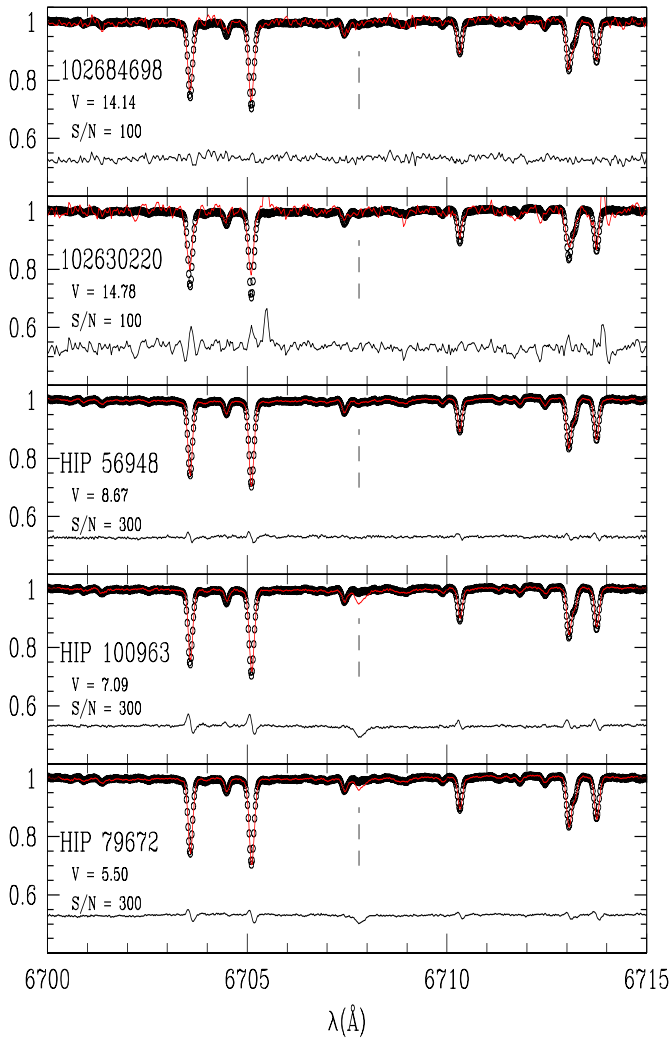}
\vspace{-0.8cm}
\caption{Region around the Li I $\lambda$6707.8~\AA~feature (marked with vertical  lines)
for the sample stars and some known solar twins. The superimposed spectrum of
the Moon (disk-integrated reflected solar spectrum) is
shown with open circles, while the stellar spectra are shown
with full red lines. Residuals (star-sun) are shown at $y=0.54$.}
\label{spectra_01} 
\end{figure}

\vspace{0.3cm}

Here we report the discovery of a new solar twin from the CoRoT database. This 
study is part of a survey of solar twins and analogs based on CoRoT and Kepler data, 
for which we are currently gathering high resolution optical spectra.  
In Sect.~\ref{sel} we describe the selection process and the observations, 
in Sect.~\ref{corottwin} we describe the analysis  and discuss  the results,  
and  in Sect.~\ref{results} we present the conclusions.

\begin{figure}[t]   
\centering
\vspace{+0.3cm}
\hspace{-0.4cm}
\includegraphics[angle=0,width=9.5cm,height=10cm]{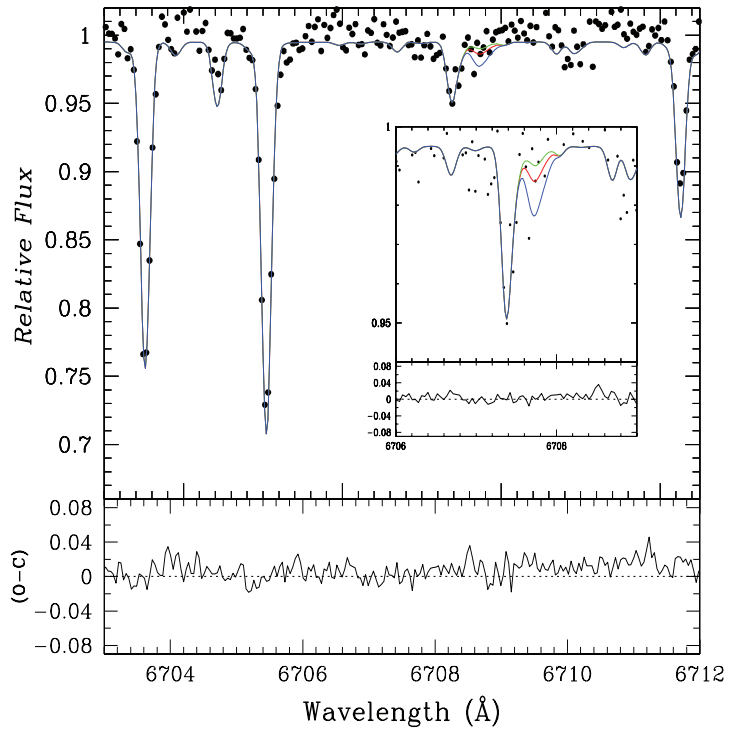}
\caption{Spectral synthesis of the Li I $\lambda$6707.8~\AA~feature of ID~102684698: 
the purely spectroscopic set of atmospheric parameters was used. 
The best-fit theoretical  profile (solid line)  and  three different values of  $A_{\rm Li}$
(0.50, 0.85, and 1.2) shown together with the observed spectrum (filled circles).  Residuals 
observed-computed (O-C) are shown on the lower panel.} 
\vspace{0.0cm}
\label{spectra_01} 
\end{figure}

\section{Selection of candidates, observations and initial analysis}  
\label{sel}
We selected solar twin candidates from a sample of more than 150,000 stars 
listed in the CoRoT \blue{\citep{auvergne2009}} database by employing precise 
2MASS photometry, as in \blue{\cite{donascimento2012}} and using the rotational modulation of
the CoRoT light curves. We also used the stellar parameters  \teff and $\log g$ 
given in \blue{\cite{sarro2013}}, and choose stars with  
$5600$~K $\leq$  \teff   $\leq 5950$~K and  $4.4 \leq \log g \leq 4.6$.  
Table~\ref{tbl:obs}  summarizes results of our chosen targets.  
We used the  available public data level 2 (N2) light curves that are ready for scientific analysis. 
These light curves were delivered by the CoRoT pipeline after nominal corrections \blue{\citep{samadi2006}}.  
To determine the $P_{\mathrm {rot}}$  for our sample,  we used the  Lomb-Scargle (LS) algorithm \blue{\citep{Scargle_1982}}.  
The light curve coverage allows us to detect reliably periods longer than 2~d 
and shorter than 50~d.  The uncertainties in   $P_{\mathrm {rot}}$  are determined by the frequency 
resolution in the power spectrum and the sampling error.  Our final sample is composed of 29 solar twins candidates. 
The derived periods and their respective errors are presented in Table~1. The $P_{\mathrm {rot}}$  uncertainty comes mainly from 
the time series limitation. To achieve the largest possible sample of solar twins with determined  $P_{\mathrm {rot}}$ ranging 
from  2 to 50 days, we analyzed all light curves  in the  CoRoT Exo-dat    \blue{\citep{Deleuil_2009}} as described by  \blue{\cite{donascimento2012}}. 

\vspace{+0.3cm}

To characterize a solar twin it is absolutely necessary to perform a detailed 
spectroscopic analysis. Hence, for our three  most promising solar twin candidates we obtained high resolution  ($R\sim60,000$) 
and high signal-to-noise ratio ($S/N\sim100$) spectra  in 2012, September 9, and 2013, March 25 (Hawaii 
Standard Time) employing  the High Dispersion Spectrograph (HDS; \blue{\citealt{noguchi2002}}) placed at the 
Nasmyth platform of  the 8.2-m Subaru Telescope. Standard data reduction  procedures (bias subtraction, 
flat-fielding, scattered-light  subtraction, spectrum extraction, wavelength calibration, continuum normalization) 
were  applied to the spectra using  IRAF\footnote{IRAF is distributed by 
the National Optical Astronomy Observatories, which is operated by the Association 
of Universities for  Research in Astronomy, Inc. under cooperative agreement with the National Science 
Foundation.}.  For stars ID~102684698 and ID~102630220 we achieved $S/N \sim$~110 and 100, respectively, 
while for  ID~110688932  the spectrum has only {$S/N\sim$ 30}, albeit for the latter it is clear 
that their lines are much broader than those in the Sun and that its Li I $\lambda$6707.8~\AA~line is much stronger. 
Thus, ID~110688932 is not a solar twin. The candidate ID~102630220 is a double-lined star, 
i.e., probably a spectroscopy binary and it was also discarded. Interestingly, a direct comparison 
of a solar spectrum (using the Moon), also observed with the same spectrograph at Subaru but 
at higher resolution ($R\sim90,000$) \blue{\citep{takeda2009}}, shows that ID~102684698 
has an overall spectrum similar to the Sun, and, more excitingly, it shows a weak Li feature, 
as in the Sun. Thus, with this model-independent analysis we determined that it is a potentially 
good solar twin. Sample spectra of the CoRoT solar twin candidates and other solar twins observed 
at Subaru \blue{\citep{takeda2009}} are shown in Figure~1.

 \begin{figure}[t]  
\centering
\vspace{-0.2cm}
\hspace{-0.5cm}
\includegraphics[angle=0,width=9.2cm,height=10.0cm]{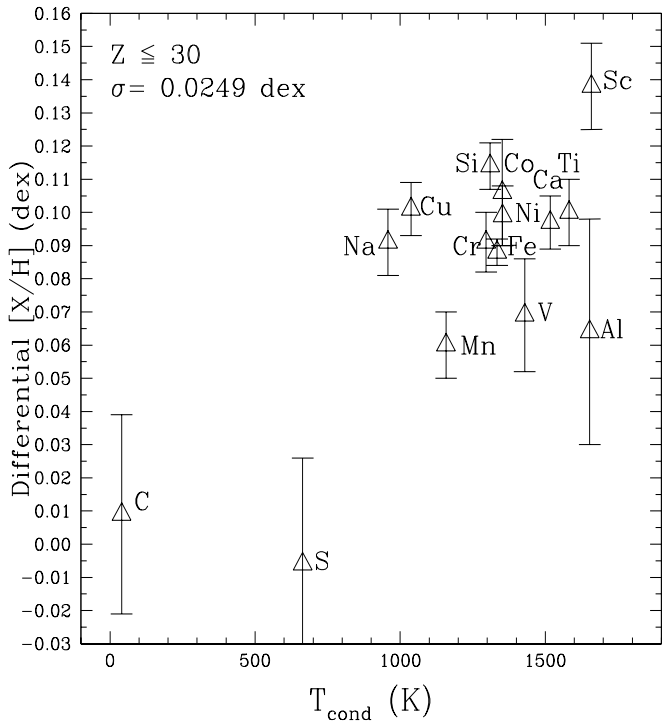}
\caption{Abundance pattern of the solar twin ID~102684698  versus 
condensation temperature. Although the error bars are relatively large 
(the element-to-element scatter from a linear fit is 0.0249 dex) due to the 
S/N (=110) of the spectrum, they are small enough to see the separation between volatiles 
and refractories.} 
\vspace{0.2cm}
\label{spectra_01} 
\end{figure}

\section{The solar twin CoRoT ID 102684698}   
\label{corottwin}

A detailed model-atmosphere analysis of ID~102684698 (CoRoT Sol 1) was performed to verify the outcome of 
the empirical comparisons and to estimate precise stellar parameters, as in \blue{\cite{ribas2010}} 
and \blue{\cite{melendez2012}}. The Li abundance was derived from 
the Li~I resonance transition at $\lambda$6707.8 \AA. A synthetic spectrum was
fitted to the Subaru spectrum for the set of atmospheric
parameters  purely spectroscopic, \teff = 5822 $\pm$ 20~K, 
$\log g$=4.31 $\pm$ 0.04, [Fe/H] = +0.09 $\pm$ 0.02 and 
$\xi$ = 1~\kms. We used  \blue{\cite{2004astro.ph..5087C}}
model-atmospheres and mostly 
laboratory $gf$-values \blue{\citep{melendez2012}} with synthetic spectra
computed using  the 2002 version of  MOOG \blue{\citep{sneden1973}}.  
\begin{figure}[h]        
\centering
\vspace{-0.2cm}
\hspace{-0.5cm}
\includegraphics[angle=0,width=9.2cm,height=10.0cm]{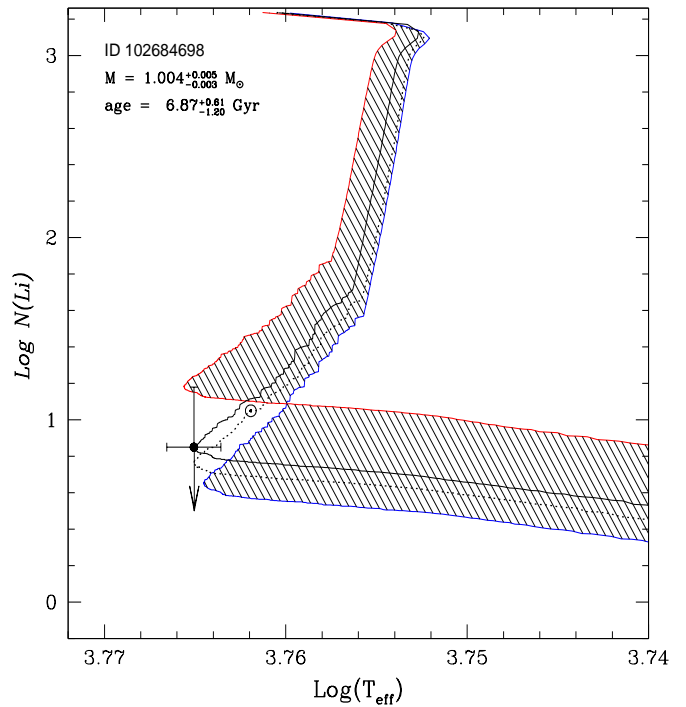}
\vspace{-0.5cm}
\caption{Li abundance for the solar twin ID~102684698 superimposed  with
the Li destruction along the evolutionary tracks as a function of the effective temperature.
The continuous line represents the 1.004 \msun~TGEC  model  passing through the
$A_{\rm Li}$ determined point.  The shaded zone represents the range of masses 
from models limited by the 1-$\sigma$ observational error bars if we  
assume $A_{\rm Li}$ = 0.85 $\pm$ 0.35 dex. [M/H] = +0.045 $\pm$ 0.015 corresponding to 
[Fe/H] = +0.09 $\pm$ 0.02 assuming [$\alpha$/H] = +0.02 $\pm$ 0.02. The position of the Sun, and the Li 
evolution of a solar model (dotted line) are also indicated.}
\label{spectra_01} 
\end{figure}
We obtained stellar parameters and chemical abundances
by differential spectroscopic equilibrium, as described in
\blue{\citep{melendez2012}}, i.e., using differential excitation
equilibrium of FeI lines to obtain \teff  and differential
ionization equilibrium of FeI and FeII to obtain $\log g$. The
surface gravity was also verified using TiI/TiII and CrI/CrII.
For the spectroscopic atmospheric parameters, we derived $A_{\rm Li}$ \ltsim~0.85  (Figure 2) in  the usual scale where 
 $A_{\rm Li}$=$\log n_{\rm Li}/n_{\rm H}+12$.  As in our previous works on solar twins, 
the analysis was performed differentially, i.e.,  the equivalent widths of the ID~102684698 
star and the Sun were measured line-by-line. 
Notice that  although the same spectrograph was used to gather both spectra, the resolving power is higher 
for the solar spectrum ($R\sim90,000$) than for the CoRoT target ($R\sim60,000$). However, all the lines 
were carefully measured by hand only after an inspection of a given line was performed both in the 
Sun and in the ID~102684698, in order to determine both the continuum and the part of the profile 
that would be used for the measurement, minimizing thus any potential systematic problem. 
The model-atmosphere analysis confirmed our empirical results, showing that indeed ID~102684698 
has stellar parameters similar to solar but with a lower $\log g$ probably showing that it is
somewhat more evolved than the Sun.  The parameters derived for the first CoRoT solar twin, ID~102684698, and our list of  
solar twins candidates, are given in Table~1.  Interestingly, ID~102684698 has an abundance pattern that 
seems different from solar and closer to that of other solar twins. This seems to expand the 
idea that the Sun has a peculiar solar composition.  Based on nearby solar twins, 
\blue{\cite{melendez2009}} and \blue{\cite{ivan2009}} showed that the Sun seems to be deficient 
in the refractory elements. As shown in Figure~3,  the Sun also shows the same deficiency of refractory elements when 
compared to ID~102684698, the farthest field solar twin so far known.

From the stellar parameters we obtained the mass and age of ID~102684698. As it seems somewhat 
more evolved than the Sun (lower $\log g$), the results from the isochrones are more reliable than 
for less evolved stars. We obtain a mass of 1.03 $\pm$ 0.03 \msun~and an age of  6.7 $\pm$ 0.6 Gyr, i.e. 
a one-solar-mass star somewhat more evolved than the Sun. That age is in the same scale 
as our previous works on solar twins, showing that our CoRoT twin is definitively older 
than the well-known solar twins as 18 Sco (2.7 $\pm$ 1.0 Gyr) or HIP 56948 (3.5 $\pm$ 0.7 Gyr)~\blue{\citep{melendez2012}}.
The mass and age of  ID~102684698  were also  estimated   by using  TGEC (Toulouse Geneva Evolution Code) 
models constrained by Li abundance (Figure~4)  as in \blue{\citet{Nascimento_2009}}.  More details of the physics 
used can be found in \blue{\citet{Richard_1996,Richard_2004}}, \blue{\citet{huibonhoa_2008}}, 
\blue{\citet{Nascimento_2009}},  and \blue{\cite{donascimento2012}}.   This   analysis agrees with a one 
solar mass twin about  2 Gyr older than the Sun.

\subsection{The lithium abundance of  CoRoT~ID~102684698}
\label{lithium}

  Stellar Li abundance is at once instructive and complex to interpret.
 The Li depletion in stars depends on several ingredients such as mass, age, metallicity,
 rotation,  magnetic fields,  mass loss,  convection treatment,  extra-mixing mechanisms 
 (e.g., \blue{\citealt{dantona1984};  \citealt{deliyannis1997}; \citealt{ventura1998};  \citealt{charbonnell1998};
 \citealt{charbonnel2005}}).  For many years, the Sun was thought to be peculiar in its 
 low Li content.  This idea was also supported  by two high-lithium,   but otherwise very 
 similar to the Sun, solar twins (HD~98168, 18~Sco)  that  have a Li abundance about three times 
higher than the Sun~\blue{\citep{melendez2006}}.  Recent studies by \blue{\cite{takeda2007}}, and \blue{\cite{melendez2007}}, 
show that this is not the case, as they have found solar twins with low Li abundances (HIP~56948 and HIP~73815).
\blue{\cite{takeda2007}}  shows that HIP~100963, is a quasi-solar twin  with higher Li abundance 
 (about 6 times solar). The spread in Li abundances among the known solar twins represents
 an opportunity to study transport mechanisms inside stars. There are already 
several models  \blue{\citep{Nascimento_2009,charbonnel2005}} that show for a fixed mass 
a depletion of Li with increasing age, which seems to be confirmed by observations in 
solar twins \blue{\citep{baumann}}.

\section{Conclusions}
\label{results}

We have found the first solar twin revealed by CoRoT and  the farthest  field solar twin so far known 
(\blue{\citealt{pasquini2008}} identified the farthest solar twins in M67), and the only solar twin older than the Sun for  which a  $P_{\mathrm {rot}}$ has been determined.  ID~102684698 (CoRoT Sol 1) has stellar parameters and mass similar to solar,  while its age {\bf is} somewhat higher than solar. Its $P_{\mathrm {rot}}$ is also slightly higher than the Sun's, i.e., consistent with an age somewhat older than solar, and its low Li abundance is also compatible with an evolved Sun. Its abundance pattern shows that the refractory elements are more enhanced than in the Sun, meaning that  our CoRoT twin is similar to that of nearby solar twins, i.e., the Sun seems to be a chemically peculiar star. We are gathering higher S/N spectra of our CoRoT solar twin and other solar twin candidates, in order to compare with better precision how typical is the Sun compared to distant solar twins and to establish the rotational evolution of the Sun. 
This  study of an unbiased sample of solar twins  with   high precision monitoring of  stellar activity cycles  for stars with well determined  $P_{\mathrm {rot}}$, 
could  better constrain models of Li depletion  as well as the gyrochronology (\blue{\citealt{barnes2003}, 2010};  
\blue{\citealt{meibom2011}, 2013}) relations.  As done for 18 Sco \blue{\citep{bazot}}, asteroseismology seems to 
be a very promising approach to classify  solar twins and can provide useful complementary information to validate 
potential solar twins candidates.

\acknowledgements
Acknowledgements to the CoRoT team. CoRoT (Convection, Rotation and planetary Transits) space 
mission, launched on 2006 December 27, was developed and is operated by the 
CNES, with participation of the Science Programs of ESA, ESA's RSSD, Austria, 
Belgium, Brazil, Germany and Spain. Based on observations obtained at the SUBARU
telescope (programs S12B-146S).  We thank the SUBARU resident astronomers and 
telescope operators for continuous  support.  JDNJr and GFPM acknowledge financial support from INCT and CNPq/Brazil.
JDNJr  acknowledges support from CNPq Universal-B 485880/2012-1.  
JM acknowledges support from FAPESP 2010/17510-3 and 2012/24392-2. 
We thank the referee for providing constructive  comments and suggestions that improved the manuscript.

\begin{table*}
\centering
\small
\caption{SUBARU  spectroscopic  observations  for the confirmed  solar twin (first part) and  
 our solar twin  candidates sample (second part).}
\label{tbl:obs}
\begin{tabular}{lccccl}
\hline
\hline
CoRoT ID    &   CoRoT  &    V$^{(\phi)}$   & \teff          &   $\log g$ &     $P_{\mathrm {rot}}$$^{(\star)}$ \\
                    &       run &        (mag) &    (K)             &           &       (d) \\
\hline
102684698   & LRa01 &	14.14 &  5822   $\pm$   20$^{(\star)}$  & 4.31$^{(\star)}$    &  29 $\pm$ 5  \\
\\[-0.4cm]
\hline
100543340	&	LRc01	&	15.22	&	5746$^{(\dagger)}$	&	4.4$^{(\dagger)}$	&	5.5 			$\pm$ 1 \\	
100567226	&	LRc01	&	15.97	&	5798$^{(\dagger)}$	&	4.6$^{(\dagger)}$	&	11.5		$\pm$ 1 \\	
100632124	&	LRc01	&	13.34	&	5633$^{(\dagger)}$	&	4.4$^{(\dagger)}$	&	21.0 		$\pm$ 2 \\	
100746852	&	LRc01	&	16.27	&	5835$^{(\dagger)}$	&	4.6$^{(\dagger)}$	&	23.9 		$\pm$ 2 \\	
100824807	&	LRc01	&	15.42	&	5751$^{(\dagger)}$	&	4.5$^{(\dagger)}$	&	14.3		$\pm$ 1 \\	
100839384	&	LRc01	&	15.41	&	5935$^{(\dagger)}$	&	4.5$^{(\dagger)}$	&	20.0 		$\pm$ 1 \\	
101030785	&	LRc01	&	14.73	&	5869$^{(\dagger)}$	&	4.6$^{(\dagger)}$	&  10.6 			$\pm$ 1 \\	
102665897	&	LRa01	&	16.55	&	5757$^{(\dagger)}$	&	4.5$^{(\dagger)}$	&   36.8 		$\pm$ 8 \\	
102709980	&	LRa01	&	14.11	&	5908$^{(\dagger)}$	&	4.5$^{(\dagger)}$	&	21.5 		$\pm$ 1 \\	
102731845	&	LRa01	&	14.06	&	5769$^{(\dagger)}$	&	4.4$^{(\dagger)}$	&	38.3 		$\pm$ 5 \\	
102739288	&	LRa01	&	16.38	&	5718$^{(\dagger)}$	&	4.4$^{(\dagger)}$	&    6.0			$\pm$ 1 \\	
102769572	&	LRa01	&	16.26	&	5750$^{(\dagger)}$	&	4.4$^{(\dagger)}$	&	4.3			$\pm$ 1 \\	
105283591	&	LRc02	&	15.25	&	5900$^{(\dagger)}$	&	4.6$^{(\dagger)}$	&	7.0 			$\pm$ 1 \\	
105398310	&	LRc02	&	14.49	&	5803$^{(\dagger)}$	&	4.5$^{(\dagger)}$	&	17.0 		$\pm$ 1 \\	
105572582	&	LRc02	&	15.56	&	5610$^{(\dagger)}$	&	4.4$^{(\dagger)}$	&	12.5 		$\pm$ 1 \\	
105595725	&	LRc02	&	12.57	&	5740$^{(\dagger)}$	&	4.5$^{(\dagger)}$	&	13.3 		$\pm$ 1 \\	
105693572	&	LRc02	&	13.01	&	5761$^{(\dagger)}$	&	4.6$^{(\dagger)}$	&	29.1 		$\pm$ 10 \\	
105735736	&	LRc02	&	14.42	&	5841$^{(\dagger)}$	&	4.6$^{(\dagger)}$	&	26 			$\pm$ 3	 \\	
105806662	&	LRc02	&	15.03	&	5610$^{(\dagger)}$	&	4.4$^{(\dagger)}$	&	15.0 		$\pm$ 1 \\	
105948587	&	LRc02	&	14.45	&	5768$^{(\dagger)}$	&	4.4$^{(\dagger)}$	&	6.5 			$\pm$ 5 \\	
106022496	&	LRc02	&	14.61	&	5641$^{(\dagger)}$	&	4.5$^{(\dagger)}$	&	12.5 		$\pm$ 1 \\	
106024409	&	LRc02	&	14.69	&	5677$^{(\dagger)}$	&	4.4$^{(\dagger)}$	&	11.2	 	$\pm$ 1 \\	
106055448	&	LRc02	&	13.57	&	5677$^{(\dagger)}$	&	4.4$^{(\dagger)}$	&	 6.1 		$\pm$ 1 \\	
110655843	&	LRa02	&	16.33	&	5641$^{(\dagger)}$	&	4.6$^{(\dagger)}$	&	8.5			$\pm$ 1 \\	
110656049	&	LRa02	&	15.1	&	5841$^{(\dagger)}$	&	4.6$^{(\dagger)}$	&	17			$\pm$ 1 \\	
110677427	&	LRa02	&	14.1	&	5878$^{(\dagger)}$	&	4.5$^{(\dagger)}$	&	$>$30  				\\	
110829375	&	LRa02	&	15.99	&	5756$^{(\dagger)}$	&	4.5$^{(\dagger)}$	&	12.0 	$\pm$ 2 \\	
110830972	&	LRa02	&	12.99	&	5749$^{(\dagger)}$	&	4.5$^{(\dagger)}$	&	2.8  	$\pm$ 1 \\[0.1cm]	
\hline
\end{tabular}
\tablecomments{$^{(\star)}$ This paper; $^{(\dagger)}$ Sarro et al. (2013); $^{(\phi)}$V~magnitude from CoRoT Exo-dat. The uncertainties in   $P_{\mathrm {rot}}$  are determined by the frequency  resolution in the power spectrum and the sampling error.  The error of our 
measurement is defined by the  probable error, which in turn is defined as 0.2865$\cdot$FWHM (full-width at half-maximum of the peak),
assuming  a Gaussian statistics around the Lomb-Scargle peak as reported  by \blue{\cite{donascimento2012}}.   }
\end{table*}

\label{obs}

\end{document}